\documentclass[twocolumn,prl,aps,showpacs,floatfix]{revtex4}

\usepackage{graphicx}%

\usepackage{dcolumn}
\usepackage{amsmath}

\newcommand{\Tf}{ T_f }

\newcommand{\ice}[1]{\relax}

\newcommand{\re}[1]{(\ref{#1})}

\newcommand{\beq}{\begin{equation}}

\newcommand{\ba}{\begin{array}}

\newcommand{\ea}{\end{array}}

\newcommand{\eeq}{\end{equation}}

\newcommand{\bea}{\begin{eqnarray}}

\newcommand{\eea}{\end{eqnarray}}

\newcommand{\nn}{\nonumber}
\newcommand{\nnb}{\nonumber}

\newcommand{\dd}{{\rm d}}

\newcommand{\sbz}{  }

\newcommand{\smsp}{  }

\newcommand{\cfB}{C_F^2 }
\newcommand{\cfAcaA}{C_F \smsp  C_A }
\newcommand{ \cfAtrAnlA}{C_F\smsp T_f}

\newcommand{\cfC}{C_F^3 }

\newcommand{\cfAtrBnlB}{ C_F \smsp T_f^2  }

\newcommand{\cfBtrAnlA}{ C_F^2\smsp T_f }

\newcommand{\cfAtrAnlAcaA}{ C_F\smsp T_f\smsp   C_A}

\newcommand{\cfBcaA}{C_F^2\smsp  C_A}

\newcommand{\cfAcaB}{C_F\smsp  C_A^2}

\newcommand{\cfD}{C_F^4 }
\newcommand{\dFFdRinvAnlA}{n_f\smsp \frac{d_F^{abcd}\smsp d_F^{abcd}}{d_R}}

\newcommand{\dFARinvA}{\frac{d_F^{abcd}\smsp d_A^{abcd}}{d_R}}

\newcommand{\cfAtrCnlC}{C_F  \smsp T_f^3}

\newcommand{\cfBtrBnlB}{C_F^2 \smsp   T_f^2}

\newcommand{\cfAtrBnlBcaA}{  C_F \smsp T_f^2\smsp C_A}

\newcommand{\cfCtrAnlA}{ C_F^3 \smsp  T_f }

\newcommand{\cfBtrAnlAcaA}{ C_F^2 \smsp T_f\smsp  C_A}

\newcommand{\cfAtrAnlAcaB}{ C_F \smsp T_f\smsp  C_A^2}

\newcommand{\cfCcaA}{C_F^3 \smsp C_A}

\newcommand{\cfBcaB}{C_F^2 \smsp C_A^2}

\newcommand{\cfAcaC}{C_F \smsp C_A^3}

\begin{document}

\title{

\ice{
{
 \vspace*{-16mm}
\centerline{\normalsize\hfill  TTP10-05}
\baselineskip 11pt
}
{}
\vspace{1mm}
}

 Adler Function, Bjorken Sum Rule, and the Crewther Relation
to  Order $\alpha_s^4$
in a  General Gauge Theory
 }     

\author{P.~A.~Baikov}
\affiliation{Skobeltsyn Institute of Nuclear Physics, Moscow State University,
Moscow~119991, Russia
        }
\author{K.~G.~Chetyrkin}\thanks{{\small Permanent address:
Institute for Nuclear Research, Russian Academy of Sciences,
 Moscow 117312, Russia}}

\author{J.~H.~K\"uhn}
\affiliation{Institut f\"ur Theoretische Teilchenphysik,
  KIT, D-76128 Karlsruhe, Germany}

\begin{abstract}

\noindent
We compute, for the first time, the order $\alpha_s^4$
contributions to the Bjorken sum rule for  polarized electron-nucleon  scattering 
and to the (non-singlet) Adler function for the  case of a  generic colour gauge group. 
We confirm at the same order a (generalized)  
Crewther relation which provides a strong test  
of the correctness of our previously obtained  results: the  QCD Adler function
and the five-loop $\beta$-function in  quenched QED.
In particular, the appearance of an  irrational contribution proportional 
to $\zeta_3$ in the latter   quantity  is confirmed.  
We obtain  the {\em commensurate} scale equation relating the  effective 
strong coupling constants as inferred from the Bjorken sum rule and from 
the Adler function at order  $\alpha_s^4$.

\end{abstract}


\maketitle

\section{Introduction}

The 
Crewther relation 
\cite{Crewther:1972kn,Broadhurst:1993ru} 
relates in a non-trivial way two seemingly  disconnected quantities, namely,
the (non-singlet) Adler \mbox{function \cite{Adler:1974gd}}  $D$  and the
coefficient function  ${C}^{Bjp}$, describing the deviation of the Bjorken sum
rule \cite{Bjorken:1969mm,Bjorken:1967px} for polarized DIS from its
naive-parton model value.
The  Adler function is defined through  the 
correlator of the vector  current $j_\mu$
\beq
3\smsp  Q^2\smsp  \Pi(Q^2) = 
i\int\dd^4 xe^{iq\cdot x}\langle 0|{\rm  T}j_\mu(x)j^\mu(0)|0\rangle
{},
\eeq
as follows 
\beq
{D}(Q^2) =  -12\, \pi^2
Q^2\, \frac{\mathrm{d}}{\mathrm{d} Q^2} \Pi(Q^2)
{},
\eeq
with $Q^2 = -q^2$. 
In fact, the Adler  function is the main theoretical object required to study
such important physical observables as the  cross section for electron-positron
annihilation into hadrons and  the hadronic  decay rates of both  the $Z$-boson and
the $\tau$-lepton (see, e.g.  \cite{Chetyrkin:1996ia}).
The Bjorken sum rule  
expresses
the integral over the  spin distributions of quarks inside of the nucleon in terms of 
its axial charge times a  coefficient function
${C}^{Bjp}$:
\bea
\Gamma_1^{p-n}(Q^2) &=& 
\int_0^1 [g_1^{ep}(x,Q^2)-g_1^{en}(x,Q^2)]dx
\nnb
\\
&=&\frac{g_A}{6}
C^{Bjp}(a_s) +
\sum_{i=2}^{\infty}\frac{\mu_{2i}^{p-n}(Q^2)}{Q^{2i-2}}
{},
\label{gBSR}
\eea
where $g_1^{ep}$ and $g_1^{en}$ are the spin-dependent proton and neutron
structure functions, $g_A$ is the nucleon axial charge as measured in 
neutron $\beta$-decay. The coefficient function
$C^{Bjp}(a_s)=1+{\cal O}(a_s)$  is proportional to the flavour-nonsinglet axial vector
current $\bar{\psi}\gamma^{\mu}\gamma_5\psi$ in the corresponding
short distance Wilson expansion. The sum  in the second line of \re{gBSR} describes for
the nonperturbative  power corrections (higher twist) which are  inaccessible for pQCD.
Within  perturbative QCD we define   
\beq
\ba{ll}
{D}(Q^2) &= d_R\left(
1 
+ \frac{3}{4}\,C_F\, a_s +
\sum_{i=2}^{\infty} \  {d}_i \, a_s^i(Q^2)
\right),
\\
{C}^{Bjp}(Q^2) &= 
1  - \frac{3}{4}\,C_F\, a_s +
\sum_{i=2}^{\infty} \  {c}_i\, a_s^i(Q^2),  
\\
1/{C}^{Bjp}(Q^2) &= 
1  + \frac{3}{4}\,C_F\, a_s +
\sum_{i=2}^{\infty} \  {b}_i\, a_s^i(Q^2)
{},
\ea
\nn
\eeq
where $d_R$ is the dimension of the quark colour representation (for
QCD $d_R=3$), $a_s \equiv \alpha_s/\pi$ and  the normalization scale
$\mu$ is set  $\mu^2=Q^2$.
Note that we consider only 
the so-called ``non-singlet''
contribition to the Adler function and do not write explictly a
common factor $\sum_i Q_i^2$ (with $Q_i$ being the  electric charge of the $i$-th quark  flavour)
for $R(s)$.

The 
Crewther relation states that 
\beq 
\ba{c} {D}(a_s)\,{C}^{Bjp}(a_s)
= d_R \left[ 1 + \frac{\beta(a_s)}{a_s}\, K(a_s) \right], 
\\ K(a_s) = 
K_0 +
a_s\,K_1 +a_s^2\,K_2 +a_s^3\,K_3 
+ \dots 
\ea \label{gCrewther} 
\eeq
Here $\beta(a_s) = \mu^2\,\frac{\mathrm{d}}{\mathrm{d} \mu^2}
a_s(\mu) =-\sum_{i \ge 0} \beta_i a_s^{i+2}$  is the QCD
$\beta$-function describing the {\em running} of the coupling constant
$a_s$ with respect to a  change of the normalization scale $\mu$ and
with its first term $ \beta_0 = \frac{11}{12}\, C_A -\frac{T}{3}\,n_f$ 
being responsible for {asymptotic freedom} of QCD. The
term proportional the $\beta$-function describes the deviation from
the limit of exact conformal invariance, with the deviations starting in order $\alpha_s^2$,  
and was
suggested \cite{Broadhurst:1993ru} on the basis of 
${\cal O}(\alpha_s^3)$ calculations of $D(a_s)$ \cite{Gorishnii:1990vf,Surguladze:1990tg} 
and ${C}^{Bjp}(a_s)$  \cite{Larin:1991tj}.
A formal proof was carried out in \cite{Crewther:1997ux,Braun:2003rp}.
The original  relation without this term was first proposed  in
\cite{Crewther:1972kn} (see, also, \cite{Adler:1973kz}).

At order $\alpha_s$ the Crewther
relation is evidently fulfilled. The colour structures
which appear in $d_n$ and $c_n$ (hence also in $b_n$) for n=1,2,3, and 4 are:
\bea
&a_s^1:& C_F, \ \ \  a_s^2: \ \   C_F^2, \, C_F \, \Tf,  C_F\,C_A,
\nnb
\label{as1}
\\
&a_s^3:&
C_F^3\, , C_F^2 \Tf\,, C_F \Tf^2\,  ,   C_F^2 C_A\, , C_F \Tf C_A  \, , C_F C_A^2,
\nnb
\label{as2}
\\
&a_s^4:&
\frac{{d_F^{a b c d} d_A^{a b c d}}}{d_R}\,
,
\,\frac{ n_f {d_F^{a b c d}d_F^{a b c d}}}{d_R} ,
C_F^4\,
,
\nnb
\label{as3}
\\
&\hspace{5mm}& 
    C_F^3 \Tf \,,
\,  C_F^2 \Tf^2 \,,
\,  C_F   \Tf^3 \,,
\,  C_F^3 C_A\,,
\,  C_F^2 \Tf  C_A,
\nnb
\\
\label{as4}
&\hspace{5mm}&
\,  C_F \Tf^2 C_A  \,, 
\,  C_F^2 C_A^2\,, 
    C_F  \Tf C_A^2\,,
\,  C_F C_A^3
{}.
\eea
Here  $C_F$ and $ C_A$ are the quadratic Casimir
operators of the 
fundamental and the adjoint representation of the Lie algebra, 
$T$ is the trace normalization of the
fundamental representation, $\Tf \equiv  T\,n_f $,  with $n_f$ being  the number of
quark flavors. The exact definitions of
${d_F^{a b c d} d_A^{a b c d}}$ and ${d_F^{a b c d}d_F^{a b c d}}$
are given in \cite{Vermaseren:1997fq}.
For  QCD (colour gauge group SU(3)):
\beq
\ba{c}
C_F =4/3\,,\, C_A=3\,,\, T=1/2\,,\, d_R = 3\,,\,
\\
{d_F^{a b c d} d_A^{a b c d}} = \frac{15}{2}\,,\, {d_F^{a b c d}d_F^{a b c d}} = \frac{5}{12}
{}.
\ea
\label{SU3}
\eeq
Note,  that all colour structures, apart  of the $d^2$-terms which appear first
at order $\alpha_s^4$, involve at least one factor  $C_F$.  
As a consequence, $K_0$ must be set to zero. An inspection of eqs. (4) and (5) 
clearly shows that the colour structures
which  may appear  in a  coefficient  $K_i$ are identical to those appearing in
the coefficient $b_{i-1}$ and $c_{i-1}$, listed in eq.~(5). 
Thus, at orders $\alpha_s^2, \alpha_s^3$ and $\alpha_s^4$ the Crewther
relation puts as many as 2, 3 and, finally, 6 constraints on the
differences $d_2 - b_2, d_3 - b_3$ and $d_4 - b_4$ respectively. The
fulfillment of these constraints constitutes   a powerful
check of the correctness of the calculations of $D^{NS}(a_s)$ and
$C^{Bjp}(a_s)$. 

Indeed, at orders ${\cal O}(\alpha_s^2)$ and ${\cal O}(\alpha_s^3)$
the results for  $D^{NS}(a_s)$ and  $1/C^{Bjp}(a_s)$
\bea
d_2 &=&
-\frac{3}{32}\cfB
{+}\cfAtrAnlA
\!
\left[
 \sbz \zeta_{3}
-\frac{11}{8}
\right]
{+}\cfAcaA
\!
\left[
\frac{123}{32}
-\frac{11\zeta_{3} }{4  } 
\right]
{},
\nn
\label{d2}
\\
b_2 &=&  
-\frac{3}{32}
\cfB
{+}\cfAtrAnlA
\!
\left[
-\frac{1}{2}\right]
{+}\cfAcaA
\!
\left[
 \frac{23}{16}\right]
{},
\nn
\label{b2}
\\
d_3 &=&  
-\frac{69}{128}
{}\cfC
{+}\cfBtrAnlA
\left[
-\frac{29}{64}
+\frac{19}{4}  \sbz \zeta_{3}
-5  \sbz \zeta_{5}
\right]
\nonumber
\\
\
{+}
\lefteqn{
\cfAtrBnlB
\left[
\frac{151}{54}
-\frac{19}{9}  \sbz \zeta_{3}
\right]
\nn
+
\cfBcaA
\left[
-\frac{127}{64}
-\frac{143}{16}  \sbz \zeta_{3}
+\frac{55}{4}  \sbz \zeta_{5}
\right]
}
\\
&{}&
\hspace{1.6cm}
+
\cfAtrAnlAcaA
\left[
-\frac{485}{27}
+\frac{112}{9}  \sbz \zeta_{3}
+\frac{5}{6}  \sbz \zeta_{5}
\right]
\nonumber
\nonumber
\\
&{}&
\hspace{1.6cm}
+
\cfAcaB
\left[
\frac{90445}{3456}
-\frac{2737}{144}  \sbz \zeta_{3}
-\frac{55}{24}  \sbz \zeta_{5}
\right]
{},
\nn
\label{d3}
\end{eqnarray}
\begin{multline*}
b_3 = 
-\frac{69}{128} \cfC
{+}\cfBtrAnlA
\left[
-\frac{299}{576}
+\frac{5}{12}  \sbz \zeta_{3}
\right]
\nonumber\\
+
\cfAtrBnlB
\left[
 \frac{115}{216}\right]
{+}\cfBcaA
\left[
\frac{1}{576}
+\frac{11}{12}  \sbz \zeta_{3}
\right]
\nonumber\\
+
\cfAtrAnlAcaA
\left[
-\frac{3535}{864}
-\frac{3}{4}  \sbz \zeta_{3}
+\frac{5}{6}  \sbz \zeta_{5}
\right]
{+}\cfAcaB
\left[
\frac{5437}{864}
-\frac{55}{24}  \sbz \zeta_{5}
\right]
{}
\label{b3}
\end{multline*}
are well consistent \cite{Broadhurst:1993ru}
with all  5 constraints on the coefficients 
$d_2, d_3, b_2$ and $b_3$ and imply 
\begin{multline*}
K_1 = C_F \left(-\frac{21}{8} + 3 \zeta_3\right) {},
\ \ \ 
K_2 = C_F \Tf\, \left(\frac{163}{24}
 - \frac{19}{3}\zeta_3\right)
\\
+ C_F C_A \,\left(-\frac{629}{32} +\frac{221}{12}\zeta_3\right)
+ C_F^2 \,\left(\frac{397}{96} + \frac{17}{2}\zeta_3 - 15 \zeta_5\right)
{}.
\end{multline*}
The next, ${\cal O}(\alpha_s^4)$, 
contribution  to $D(a_s)$ has been recently computed  \cite{Baikov:2008jh}
for QCD, i.e.  setting  the colour structures to their $SU(3)$ numerical values  (eq.~\re{SU3}).
The function   $C^{Bjp}(a_s)$ is known to order 
$\alpha_s^3$  {\em only}.

The importance of computation of the ${\cal O}(\alpha_s^4) $
contribution to the both coefficients $d_4$ and $b_4$ for a
generic colour gauge group comes from a few reasons.

First, the knowledge of $c_4$ in the Bjorken sum rule  is vital for proper
extraction of higher twist contributions. Indeed, in
\cite{Pasechnik:2008th} the recent Jefferson Lab  data on the spin-dependent
proton and neutron structure functions 
\cite{Bosted:2006gp,Prok:2008ev,Deur:2008ej,Deur:2004ti,Deur:2005cf}
were used to extract
the leading and subleadinrg higher twist parameters $\mu_4$ and $\mu_6$. It has been
demonstrated that, say, the twist four term $\mu_4$ approximately
halves its value in transition from LO to NLO, and from NLO to NNLO.
This  {\em duality} between perturbative and
non-perturbative contributions has  been observed before for the
structure function $F_3$ \cite{Kataev:1999bp} (for a related recent
discussion see also \cite{Narison:2009ag}).

Second, the Bjorken sum rule provides us with a very convenient definition of the
{\em effective strong coupling constant} (ECC)  \cite{Grunberg:1980ja,Deur:2005cf}, namely,
\beq
6\,\Gamma_1^{p-n}(Q^2) = 
g_A\,\left(1- a_{g_1}(Q^2)\right)
\label{ECC_g1}
{}.
\eeq
This quantity is directly measurable down to vanishing values of $Q^2$
and, due to eq.~\re{gBSR}, approaches to the standard $\alpha_s(Q)$ at
large $Q^2$. It is by definition gauge and scheme invariant.  %
Another convenient ECC, $a_{D}$, comes from the Adler function \cite{Brodsky:2002nb}:
\beq
D(Q^2) = 
1 + a_{D}(Q^2)
{}.
\label{ECC_D}
\eeq
As its perturbative expansion is available
to ${\cal O}(\alpha_s^4)$  \cite{Baikov:2008jh} the knowledge of $c_4$
will allow for the first time to compare two ECC's with the help of a
commensurate scale relation \cite{Brodsky:1994eh} at an order unprecedented to date.

\newcommand{\qQED}{{\rm qQED}}

Third, the six constraints imposed by eq.~\re{gCrewther} provide a
highly nontrivial and welcome check of the calculation of $d_4$ in QCD
\cite{Baikov:2008jh}. In particular, in \cite{Baikov:2008cp} we
computed a part of the full result for $d_4$, namely, the term
proportional to the colour structure $C_F^4$.  As is well-known, an
interesting object -- the $\beta$-function of  quenched QED --- can
be inferred from the part of the Adler function which depends on 
$C_F$  only by setting $C_F=1$ and adjusting a global normalization factor.
The result ($A\equiv \frac{\alpha}{4\pi}$) 
\beq
\ba{ll}
{\bf \beta}^{\qQED} &=
\frac{4}{3}\, A
 + 4\,A^2 -{2}\,A^3 -46  A^4 
\\
&+
\left(\frac{4157}{6}  \ \  
+ \ \ 128\,  \zeta_3\right)\, A^5
\label{betaqQED_5l}
\ea
\eeq
revealed an unexpected \footnote{
Unexpected, because there existed a wide-spread belief that the
rationality property is not accidental but holds also in higher orders
\cite{Bender:1976pw,Broadhurst:1999xk}.}  appearance of the irrational constant
$\zeta_3$ at five loops and had  cast  doubt on  the
correctness of the full QCD result for $d_4$ \cite{Kataev:2008sk}.

Using the same techniques  as in calculations of  \cite{Baikov:2008jh} and \cite{Larin:1991tj}
we have computed
the  the Adler function and the function  ${C}^{Bjp}$ for a general gauge  group to order 
$\alpha_s^4$. Our results read 
\begin{widetext}
\begin{eqnarray}
d_4 & = &\dFARinvA
\left[
\frac{3}{16} 
-\frac{1}{4}  \sbz \zeta_{3}
-\frac{5}{4}  \sbz \zeta_{5}
\right]
{+}\dFFdRinvAnlA
\left[
-\frac{13}{16} 
-  \sbz \zeta_{3}
+\frac{5}{2}  \sbz \zeta_{5}
\right]
{+}\cfD
\left[
\frac{4157}{2048} 
+\frac{3}{8}  \sbz \zeta_{3}
\right]
\nonumber\\
&{+}&\cfCtrAnlA
\left[
\frac{1001}{384} 
+\frac{99}{32}  \sbz \zeta_{3}
-\frac{125}{4}  \sbz \zeta_{5}
+\frac{105}{4}  \sbz \zeta_{7}
\right]
{+}\cfBtrBnlB
\left[
\frac{5713}{1728} 
-\frac{581}{24}  \sbz \zeta_{3}
+\frac{125}{6}  \sbz \zeta_{5}
+3  \sbz \zeta_3^2
\right]
{+}\cfAtrCnlC
\left[
-\frac{6131}{972} 
+\frac{203}{54}  \sbz \zeta_{3}
+\frac{5}{3}  \sbz \zeta_{5}
\right]
\nn
\\
&{+}&\cfCcaA
\left[
-\frac{253}{32} 
-\frac{139}{128}  \sbz \zeta_{3}
+\frac{2255}{32}  \sbz \zeta_{5}
-\frac{1155}{16}  \sbz \zeta_{7}
\right]
{+}\cfBtrAnlAcaA
\left[
\frac{32357}{13824} 
+\frac{10661}{96}  \sbz \zeta_{3}
-\frac{5155}{48}  \sbz \zeta_{5}
-\frac{33}{4}  \sbz \zeta_3^2
-\frac{105}{8}  \sbz \zeta_{7}
\right]
\nonumber\\
&{+}&\cfAtrBnlBcaA
\left[
\frac{340843}{5184} 
-\frac{10453}{288}  \sbz \zeta_{3}
-\frac{170}{9}  \sbz \zeta_{5}
-\frac{1}{2}  \sbz \zeta_3^2
\right]
{+}\cfBcaB
\left[
-\frac{592141}{18432} 
-\frac{43925}{384}  \sbz \zeta_{3}
+\frac{6505}{48}  \sbz \zeta_{5}
+\frac{1155}{32}  \sbz \zeta_{7}
\right]
\nonumber\\
&{+}&\cfAtrAnlAcaB
\left[
-\frac{4379861}{20736} 
+\frac{8609}{72}  \sbz \zeta_{3}
+\frac{18805}{288}  \sbz \zeta_{5}
-\frac{11}{2}  \sbz \zeta_3^2
+\frac{35}{16}  \sbz \zeta_{7}
\right]
\nonumber\\
&{+}&\cfAcaC
\left[
\frac{52207039}{248832} 
-\frac{456223}{3456}  \sbz \zeta_{3}
-\frac{77995}{1152}  \sbz \zeta_{5}
+\frac{605}{32}  \sbz \zeta_3^2
-\frac{385}{64}  \sbz \zeta_{7}
\right]
{},
\label{d4}
\end{eqnarray}
\begin{eqnarray}
b_4 & = &\dFARinvA
\left[
\frac{3}{16} 
-\frac{1}{4}  \sbz \zeta_{3}
-\frac{5}{4}  \sbz \zeta_{5}
\right]
{+}\dFFdRinvAnlA
\left[
-\frac{13}{16} 
-  \sbz \zeta_{3}
+\frac{5}{2}  \sbz \zeta_{5}
\right]
{+}\cfD
\left[
\frac{4157}{2048} 
+\frac{3}{8}  \sbz \zeta_{3}
\right]
\nonumber\\
&{+}&\cfCtrAnlA
\left[
-\frac{473}{2304} 
-\frac{391}{96}  \sbz \zeta_{3}
+\frac{145}{24}  \sbz \zeta_{5}
\right]
{+}\cfBtrBnlB
\left[
\frac{869}{576} 
-\frac{29}{24}  \sbz \zeta_{3}
\right]
{+}\cfAtrCnlC
\left[
-\frac{605}{972}\right]
\nonumber\\
&{+}&\cfCcaA
\left[
-\frac{8701}{4608} 
+\frac{1103}{96}  \sbz \zeta_{3}
-\frac{1045}{48}  \sbz \zeta_{5}
\right]
{+}\cfBtrAnlAcaA
\left[
-\frac{17309}{13824} 
+\frac{1127}{144}  \sbz \zeta_{3}
-\frac{95}{144}  \sbz \zeta_{5}
-\frac{35}{4}  \sbz \zeta_{7}
\right]
\nonumber\\
&{+}&\cfAtrBnlBcaA
\left[
\frac{165283}{20736} 
+\frac{43}{144}  \sbz \zeta_{3}
-\frac{5}{12}  \sbz \zeta_{5}
+\frac{1}{6}  \sbz \zeta_3^2
\right] 
{+}\cfBcaB
\left[
-\frac{435425}{55296} 
-\frac{1591}{144}  \sbz \zeta_{3}
+\frac{55}{9}  \sbz \zeta_{5}
+\frac{385}{16}  \sbz \zeta_{7}
\right]
\label{b4}
\\
\ \ 
\nonumber
{+}\cfAtrAnlAcaB
&{}&
\!\!\!\!\!\!\!\!\!\!\!\!
\left[
-\frac{1238827}{41472} 
-\frac{59}{64}  \sbz \zeta_{3}
+\frac{1855}{288}  \sbz \zeta_{5}
-\frac{11}{12}  \sbz \zeta_3^2
+\frac{35}{16}  \sbz \zeta_{7}
\right]
+\cfAcaC
\left[
\frac{8004277}{248832} 
-\frac{1069}{576}  \sbz \zeta_{3}
-\frac{12545}{1152}  \sbz \zeta_{5}
+\frac{121}{96}  \sbz \zeta_3^2
-\frac{385}{64}  \sbz \zeta_{7}
\right]
{}.
\end{eqnarray}
All six constraints from the generalized Crewther relation  are indeed met with
\bea
K_3  &=& 
 C_F^3 
\left(
\frac{2471}{768} + \frac{61}{8}\zeta_3 - \frac{715}{8}\zeta_5 +
\frac{315}{4}\zeta_7
\right)
+
C_F^2 \Tf \left(-\frac{7729}{1152} - \frac{917}{16}\zeta_3
+\frac{125}{2}\zeta_5+ 9 \zeta_3^2 \right)
\nonumber\\
&+& 
C_F \Tf^2  \left(-\frac{307}{18} + \frac{203}{18}\zeta_3 + 5 \zeta_5\right) 
+
C_F^2 C_A  \left(
\frac{99757}{2304} + \frac{8285}{96}\zeta_3 - \frac{1555}{12}\zeta_5 -
\frac{105}{8}\zeta_7
\right)
 \nonumber
\\
&+&  C_F \Tf   C_A 
\left(\frac{1055}{9} - \frac{2521}{36}\zeta_3
-\frac{125}{3}\zeta_5- 2 \zeta_3^2 
\right)
+ C_F C_A^2 \left(
-\frac{406043}{2304} + \frac{18007}{144}\zeta_3
+ \frac{2975}{48}\zeta_5- \frac{77}{4}\zeta_3^2
\right)
{}.
\nonumber
\eea
\end{widetext}
Note, that   coefficients in front  of first three colour structures in 
eqs.~(\ref{d4},\ref{b4}) ($C_F^4, \dFFdRinvAnlA$ and $\dFARinvA$)
are  equal,  as they should. The $C_F^4$-term, in  particular, provides us  with a beautiful
confirmation of the correctness of the result \re{betaqQED_5l} for  the  qQED 
$\beta$-function (the test was originally suggested in \cite{Kataev:2008sk}).

It is interesting to note that the results do not depend on $\zeta_n$ with 
$n=2,4,6$.
Also, unexpected feature of our results is the {\em separate}
proportionality all terms of highest and sub-highest transcendentality in 
a given loop order (that is  $\zeta_3^2$ and $\zeta_7$ at $\alpha_s^4$, 
$\zeta_5$ at $\alpha_s^3$ and,
at last, $\zeta_3$ at $\alpha_s^2$) to $\beta_0$. 
This feature {\em is
not} required by \re{gCrewther}, the latter essentially
constraints only the {\em difference} $d_i - b_i$.


In numerical form  $C^{Bjp}$ reads (with all colour factors set
to their QCD values)
\begin{eqnarray}
{}C^{Bjp} &=&
1
{-} a_s
{+} 
\left(
-4.583
+0.3333  \,n_f
\right)a_s^2
\\
&{+}& a_s^3
\left(
-41.44
+7.607  \,n_f
-0.1775  \, n_f^2
\right) a_s^3
\nonumber\\
&{+}& 
\left(
-479.4
+123.4  \,n_f
-7.697  \, n_f^2
+0.1037  \, n_f^3
\right)\,a_s^4
{}.
\nonumber
\label{CBJN}
\end{eqnarray}
It is of interest to compare  the newly found  coefficient in front
of the $\alpha_s^4$  term with well-known predictions \cite{Kataev:1995vh}
\[c_4^{\rm pred} (n_f=3,4,5,6) = -130,\, -58,\, -18,\, 22 \]
and 
\[c_4^{\rm exact}(n_f=3,4,5,6) = 
-175.7, \, -102.4, \, -41.96, \, 6.2
{}.
\]


\newcommand{\QstarS}{{Q}^{\star 2}}

At last, we derive the commensurate relation connecting two the ECC's
$a_{g_1}$ and $a_{D}$ as defined in eqs.~(\ref{ECC_g1},\ref{ECC_D}).
Following  ref.~\cite{Brodsky:1995tb} we get for QCD 
\beq
\left(
1+a_{D}(\QstarS)
\right)
\left(
1-
a_{g_1}(Q^2)
\right) =1
\label{commmens}
{},
\eeq
with ($a_D^\star = a_D(\QstarS)$)
\begin{multline*}
\ln\left(\frac{\QstarS}{Q^2}\right)
= -K_1 + a_D^\star\left[
\rule{0mm}{3.4mm}
  \beta _0 \, K_1^2+2 d_2 \, K_1-\, K_1-\, K_2
\rule{0mm}{3.4mm}
\right]
\\
{}+ \left(a_D^\star\right)^2
\left[
\rule{0mm}{3.4mm}
\beta _0 \left(-6 d_2 \, K_1^2+2 \, K_1^2+3 \, K_2 \, K_1\right)
-2 \beta _0^2 \, K_1^3
\right.
\\
{}
\left.
+K_1\left(
\frac{3}{2} \beta _1 K_1
-6 d_2^2 +2 d_2 +3 d_3
\right)
+
K_2\left(
3 d_2  - 1
\right)
-K_3
\rule{0mm}{3.4mm}
\right]
{}
\\
=
-1.30823 + a_D^\star\left[0.80241 - 0.03933\,n_f\right]
\\
+
\left(a_D^\star\right)^2
\left[ -16.9020 + 2.62311 n_f - 0.10202 n^2_f     \right]
{}.
\end{multline*}


In conclusion we want to mention that all our calculations have been
performed on a SGI ALTIX 24-node IB-interconnected cluster of 8-cores
Xeon computers and on the HP XC4000 supercomputer of the federal state
Baden-W\"urttemberg
using  parallel  \cite{Tentyukov:2004hz} as well as thread-based \cite{Tentyukov:2007mu} versions  of FORM
\cite{Vermaseren:2000nd}.  For evaluation of color factors we have used the FORM program {\em COLOR}
\cite{vanRitbergen:1998pn}. The diagrams have been generated with QGRAF \cite{Nogueira:1991ex}.

This work was supported by the Deutsche Forschungsgemeinschaft in the
Sonderforschungsbereich/Transregio SFB/TR-9 ``Computational Particle
Physics'' and  by RFBR grant 08-02-01451. We thank V.~M.~Braun for useful discussions.

\end{document}